\documentstyle[prl,aps]{revtex}

\def\la{\langle}
\def\ra{\rangle}

\def\lb{\lbrack}
\def\rb{\rbrack}

 \def\Slash#1{
  \begin{picture}(5,6)(0,0)
  \put(-.7,-1.2){\line(5,6)6}
  \end{picture}
  \kern-.8em#1}
 \def\slash#1{
  \begin{picture}(5,6)(0,0)
  \put(-1.5,-1.7){\line(5,6)5}
  \end{picture}
  \kern-.8em#1}

\def\Sn{\Slash \nabla}

\def\gg5{\gamma_5}
\def\hg5{\hat{\gamma}_5}
\def\g4{\gamma_4}

\def\O{{\cal O}}

\def\U{{\cal U}}

\def\V{{\cal V}}

\def\hD{{\textstyle \frac{1}{2}}\Delta}

\def\Qlatmr1{Q_{lat}^{(m=r=1)}}

\def\be{\begin{eqnarray}}
\def\ee{\end{eqnarray}}

\def\Nb{N_{\beta}}
\def\bx{{\bf x}}
\def\bp{{\bf p}}
\def\t{\tau}
\def\hpsi{\hat{\Psi}}
\def\hU{\hat{U}}
\def\hD{\hat{D}}
\def\hM{\hat{M}}
\def\hU{\hat{U}}
\def\hcU{\hat{{\cal U}}}

\def\hV{\hat{{\cal V}}}
\def\hL{\hat{L}}
\def\hJ{\hat{J}}
\def\hK{\hat{K}}

\begin{document}

\draft
 
\title{A simplified test of universality in Lattice QCD}

\author{David H. Adams}

\address{Instituut-Lorentz for Theoretical Physics, Leiden University, 
Niels Bohrweg 2, NL-2333 CA Leiden, The Netherlands. Email: adams@lorentz.leidenuniv.nl}

\date{16 Dec.'03}

\maketitle

\begin{abstract}

A simplified test of universality in Lattice QCD is performed by
analytically evaluating the continuous Euclidean time limits of various lattice 
fermion determinants, both with and without a Wilson term to lift the fermion doubling
on the Euclidean time axis, and comparing them with each other and with 
zeta-regularised fermion determinant in the continuous time---lattice space setting. 
The determinant relations expected from universality considerations are found to be
violated by a certain gauge field-dependent factor, i.e. we uncover a ``universality
anomaly''. The physical significance, or lack thereof, of this factor is a delicate
question which remains to be settled.

\end{abstract}


\widetext

The low-lying spectrum of the naive lattice Dirac operator approximates the low-lying 
spectrum of the continuum Dirac operator but with a 16-fold degeneracy due to
fermion doubling \cite{Wilson}. Lattice QCD with a naive fermion is therefore regarded
as a regularisation of continuum QCD with 16 degenerate fermion flavours. Lattice QCD
with a staggered fermion \cite{staggered} is regarded as a regularisation of 4 flavour QCD,
in accordance with the fact that one naive fermion flavour is equivalent after
spin diagonalisation to 4 staggered fermion flavours \cite{diag}. Lattice QCD with a 
Wilson fermion \cite{Wilson}, where the Wilson term is added to the naive fermion 
action to lift the fermion doubling, is regarded as a regularisation of QCD with
a single fermion flavour. Implicit in this is a {\em universality hypothesis}:
the LQCD's with naive fermion, staggered fermion, and Wilson fermion
are all in the right universality class to reproduce continuum QCD, the only 
difference being in the number of continuum fermion flavours the different lattice 
fermion formulations describe.

It is highly desirable to test this universality hypothesis wherever possible.
This is particularly important in view of the fact that LQCD calculations
with both dynamical Wilson and staggered fermions are currently being pursued at 
great effort and expense (see \cite{Jansen} for a recent review with references
to the literature). An interesting quantity to consider in this context is the 
fermion determinant, which appears in the QCD functional integral when the fermions 
are dynamical. In LQCD with dynamical staggered fermions, the 
fourth root of the staggered fermion determinant is used as the fermion determinant
for a single quark flavour. An important test of the universality
hypothesis would therefore be to check whether the fourth power of the Wilson
fermion determinant
coincides with the staggered fermion determinant in the continuum limit, or,
equivalently, whether the 16th power of the Wilson determinant coincides with
the naive fermion determinant in this limit. Such a test appears analytically
impossible with currently known techniques though. However, a simplified version
of this test is feasible: instead of the full continuum limit one can take the 
continuum limit for one of the spacetime coordinates while keeping the remaining
coordinates discrete. The effect of including in the fermion action a Wilson-type term
to lift the fermion doubling along the chosen coordinate axis can then be investigated.
In this paper we perform such a simplified test of universality
by evaluating the continuous Euclidean time limits of the various lattice fermion 
determinants, both with and without the time part of the Wilson term in the action, 
and comparing them with each other and with the fermion 
determinant in the continuous time---lattice space setting. (The fermion determinant
in the latter setting is defined via zeta-regularisation.)

On a finite spacetime lattice, with $N_{\sigma}$ sites along the spacial axes
$\sigma=1,2,3\,$, $\,\Nb$ sites along the Euclidean time axis, time
lattice spacing=$a$, spacial lattice spacing=$a'$, setting
$\beta=a\Nb$=time length (=inverse temperature in the
finite temperature QCD setting), we consider lattice fermion actions of the form
$S=a(a')^3\sum_{(\bx,\t)}\bar{\psi}(\bx,\t)D^{(r)}\psi(\bx,\t)$
where $(\bx,\t)$ runs over the lattice sites and 
\be
D^{(r)}&=&\g4{\textstyle \frac{1}{a}}\nabla_4+{\textstyle \frac{r}{2a}}\Delta_4
+D_{space}+m \label{2} \\
\nabla_4\psi(\bx,\t)&=&{\textstyle \frac{1}{2}}(U_4(\bx,\t)\psi(\bx,\t+a)
-U_4(\bx,\t-a)^{-1}\psi(\bx,\t-a)) \label{3} \\
\Delta_4\psi(\bx,\t)&=&2\psi(\bx,\t)-U_4(\bx,\t)\psi(\bx,\t+a)
-U_4(\bx,\t-a)^{-1}\psi(\bx,\t-a) \label{4} \\
D_{space}^{(r')}&=&{\textstyle \frac{1}{a'}}\Sn_{space}
+{\textstyle \frac{r'}{2a'}}\Delta_{space} \label{5} \\
\Sn_{space}\psi(\bx,\t)&=&\sum_{\sigma=1}^3\gamma_{\sigma}{\textstyle \frac{1}{2}}
(U_{\sigma}(\bx,\t)\psi(\bx+\hat{\sigma},\t)
-U_{\sigma}(\bx-\hat{\sigma},\t)^{-1}\psi(\bx-\hat{\sigma},\t)) \label{6} \\
\Delta_{space}\psi(\bx,\t)&=&\sum_{\sigma=1}^3
(2\psi(\bx,\t)-U_{\sigma}(\bx,\t)\psi(\bx+\hat{\sigma},\t)
-U_{\sigma}(\bx-\hat{\sigma},\t)^{-1}\psi(\bx-\hat{\sigma},\t))\qquad\  \label{7}
\ee
For $r=r'=0$ this is the naive fermion action while for $r=r'\ne0$ it is the Wilson
action. We are going to evaluate the continuous time limit ($a\to0\,$, $\Nb\to\infty$
with $\beta=a\Nb$ held fixed) of the fermion determinants $detD_{\alpha}^{(r)}$
at $r=0$ and $r=1$ and compare them with each other and
with the zeta-regularised fermion determinant $det_{\zeta}D_{\alpha}$
in the continuous time---lattice space setting. 
The Dirac operator in the latter setting is 
\be
D=\g4({\textstyle \frac{d}{d\t}}+A_4)+D_{space}+m
\label{8}
\ee
with $A_4(\bx,\t)$ being the 4-component of a smooth continuum gauge field such that
$U_4(\bx,\t)=T\,e^{\int_0^1aA_4(\bx,\t+(1-t)a)dt}$
(T=t-ordering) is the lattice transcript. The subscripts ``$\alpha$'' in 
$D_{\alpha}^{(r)}$ and $D_{\alpha}$ refer to the operators defined by replacing 
$U_4\to e^{-\alpha a}U_4$ in (\ref{2})--(\ref{7}) and $A_4\to A_4-\alpha$ in
(\ref{8}), respectively. The role of the complex parameter $\alpha$ is to incorporate
the effect of a general boundary condition at the time boundaries: 
$D_{\alpha}^{(r)}$ (resp. $D_{\alpha}$) with {\em periodic} time b.c. has the same
spectrum and determinant as $D^{(r)}$ (resp. $D$) with time b.c.
$\psi(\bx,\beta)=e^{\alpha\beta}\psi(\bx,0)$.
Thus the introduction of $\alpha$ allows us to always take periodic time b.c. when considering
the fermion determinant. It can also be used to incorporate a chemical 
potential $\mu\,$: QCD at finite temperature and density, where the fermion fields 
satisfy anti-periodic time b.c., corresponds to $\alpha=\mu+i\pi/\beta$. 
The gauge fields are required to satisfy periodic time b.c. The spacial b.c.'s for the
fermion and gauge fields will not play a role in our considerations; we can simply take them
to be periodic.

The term $\frac{r}{2a}\Delta_4$ in (\ref{2}) is the ``time part'' of the usual Wilson term.
It lifts the fermion doubling on the Euclidean time axis when $r\ne0$. Therefore, if we
think of the continuous time---lattice space setting as the ``continuum setting'', then the
aforementioned universality hypothesis, which relates the continuum limits of the naive, 
staggered and Wilson fermion determinants, translates into 
the following \hfill\break
{\em Simplified universality hypothesis}:
\be
\lim_{a\to0}\ detD_{\alpha}^{(0)}=\Big(\,\lim_{a\to0}\ detD_{\alpha}^{(r)}
\Big)_{r\ne0}^2\qquad\ \mbox{(mod p.i.f.'s)}
\label{10}
\ee
where ``p.i.f.'s'' refers to 'physically inconsequential factors'. 
This is now something
which can be checked analytically. Our main technical result in this paper is 
the following \hfill\break
{\em Result of calculation}:
\be
\lim_{a\to0}\ detD_{\alpha}^{(0)}=\Big(\,\lim_{a\to0}\ detD_{\alpha}^{(1)}
\Big)^2\,e^{-\int_0^{\beta}Tr\,(\frac{r'}{2a'}\Delta_{space}(\t))\,d\t}
\qquad\ \mbox{(mod p.i.f.'s)} \label{12}
\ee
where $\Delta_{space}(\t)$ is defined on the space of lattice spinor fields $\psi(\bx)$,
living only on the spacial lattice, by replacing $\psi(\bx,\t)$ by $\psi(\bx)$ in
(\ref{7}).
The p.i.f.'s in (\ref{12}) are gauge field-independent factors whose only effects
are to produce constant (vacuum) shifts in certain physical quantities. They include
inverse powers of $a$ which diverge in the $a\to0$ limit.

The result (\ref{12}) reveals a ``universality anomaly'': the exponential factor in the 
right-hand side violates the universality expectation (\ref{10}).
Thus it is important to ascertain the significance, or lack thereof, of this factor. 
Since it is gauge field-dependent it cannot strictly speaking be regarded as
a p.i.f. in the continuous time---lattice space theory. However, since the spacial
Wilson term $\frac{r'}{2a'}\Delta_{space}$ formally vanishes in the spacial continuum limit,
one could argue that the exponential factor is effectively a p.i.f. when one goes on to take 
that limit. This is a delicate issue though, since 
$Tr\,\frac{1}{a'}\Delta_{space}$ actually diverges in in the $a'\to0$ limit (note that
in the free field case the largest eigenvalue of $\frac{1}{a'}\Delta_{space}$ is
$\sim\frac{1}{a'}$). Further study is required to clarify this issue.

Since the reasoning which leads to the universality expectation (\ref{10}) is the same 
as that which leads to the expectation that LQCD with naive, staggered and Wilson
fermions are all in the same universality class, (\ref{12}) is a potential reason 
for concern about whether the latter universality hypothesis really holds. It would therefore
be a significant reassurance if the anomaly factor in (\ref{12}) can be shown to be
physically inconsequential. It should be 
noted, however, that even if this turns out not to be the case, 
it would not in itself invalidate the universality hypothesis
for LQCD with naive, staggered and Wilson fermions, since the comparison between naive fermion 
($r=r'=0$) and Wilson fermion ($r=r'\ne0$) is not covered by (\ref{12}) and we could still 
be lucky in this case. But it would certainly raise a serious concern.

Remarkably, the anomaly in (\ref{12}) is mirrored by an ambiguity
in the zeta-regularised fermion determinant in the continuous time---lattice space
setting. The latter can be expressed either as $det_{\zeta}D_{\alpha}$ or
$det_{\zeta}(\gamma_4D_{\alpha})$ (since $\bar{\psi}=\psi^*\gamma_4$). Formally, the 
determinants of $D_{\alpha}$ and $\gamma_4D_{\alpha}$ coincide, but it turns out
that the rigorously defined zeta-determinants do not. In fact we find
$det_{\zeta}(\gamma_4D_{\alpha})=
e^{-\frac{1}{2}\int_0^{\beta}Tr\,(\frac{r'}{2a'}\Delta_{space}(\t))
\,d\t}\,det_{\zeta}D_{\alpha}$ (mod p.i.f.'s), and
\be
\lim_{a\to0}\ detD_{\alpha}^{(1)}=det_{\zeta}D_{\alpha}\quad,\quad\
\lim_{a\to0}\ detD_{\alpha}^{(0)}=det_{\zeta}(\gamma_4D_{\alpha})^2
\qquad\mbox{(mod p.i.f.'s)} \label{13}
\ee
This shows that the lattice regularisations are consistent with zeta-regularisations of the 
fermion determinant in the continuous time---lattice space setting, and that the requirement
that the anomaly factor in (\ref{12}) be physically inconsequential is also necessary
for consistency of continuous time--lattice space QCD when the fermion determinant is defined
by zeta-regularisation.

In the remainder of this paper we sketch the derivation of (\ref{12})--(\ref{13}) and 
give other, more explicit, expressions for the $a\to0$ limits of $detD_{\alpha}^{(0)}$ and
$detD_{\alpha}^{(1)}$. The full details are provided in \cite{DA(det)}. 
It is convenient to regard $\psi(\bx,\t)$ as a function $\Psi(\t)$ living on the lattice 
sites of the Euclidean time axis and taking values in the vectorspace $W=\{\psi(\bx)\}$, 
i.e. the space of lattice spinor fields living on the spacial lattice only.
Set $N:=\dim W$. 
Define the linear operator $U_4(\t)$ on $W$ by $(U_4(\t)\psi)(\bx)=U_4(\bx,\t)\psi(\bx)$.
The operator $D_{space}(\t)$ on $W$ is defined similarly by replacing
$\psi(\bx,\t)$ by $\psi(\bx)$ in (\ref{5})--(\ref{7}). 
Since $\Psi(\beta)=\Psi(0)$
we can represent $\Psi$ by the vector $\hpsi=(\hpsi(0),\dots,\hpsi(N_{\beta}-1))$ where
$\hpsi(k)=\Psi(ka)$. Then $D^{(r)}$ is represented by
\be
\hD^{(r)}\hpsi(k)=d_{-1}^{(r)}(k)\hpsi(k-1)+d_0^{(r)}(k)\hpsi(k)+d_1^{(r)}(k)\hpsi(k+1)
\label{15}
\ee
where the operators $d_j^{(r)}(k):W\to W$ are given by
$d_1^{(r)}(k)=\frac{1}{2a}\,(\g4-r)\,\hU_4(k)\,$,
$\;d_{-1}^{(r)}(k)=-\frac{1}{2a}\,(\g4+r)\,\,\hU_4(k-1)^{-1}\,$, 
$\;d_0^{(r)}(k)=\frac{r}{a}+\hD_{space}(k)+m$
with $\hU_4(k):=U_4(ka)$ and $\hD_{space}(k):=D_{space}(ka)$. 
The generalisation of $\hD^{(r)}$
to $\hD_{\alpha}^{(r)}$, given by $U_4\to e^{-\alpha a}U_4\,$,
is equivalent to $d_{\pm1}^{(r)}\to e^{\mp \alpha a}d_{\pm1}^{(r)}$ in (\ref{15}). 
After writing $\hD_{\alpha}^{(r)}$ as an $\Nb\times\Nb$ matrix, its determinant can be
straightforwardly evaluated via the method of \cite{Gibbs}. The cases $r=1$ and $r\ne1$ 
require separate treatments due to the fact that $d_{\pm1}^{(r)}(k)$ is invertible when
$r\ne1$ but not when $r=1$. The details of the calculation are provided in \cite{DA(det)};
here we simply quote the results, assuming for convenience that $\Nb$ is even in the 
$r\ne1$ case:
\be
detD_{\alpha}^{(r\ne1)}&=&\Big(\frac{(1-r^2)^2}{2a}\Big)^{N\Nb}\,e^{-\alpha\beta N}
\,det\!\left(\left({{\bf 1} \atop 0}\;{0 \atop {\bf 1}}\right)
-e^{\alpha\beta}\,\hcU^{(r)}(\Nb/2)\right) \label{22} \\
detD_{\alpha}^{(1)}&=&\Big(\frac{1}{a}\Big)^{N\Nb}\,e^{-\alpha\beta N/2}
\,\bigg\lb\prod_{k=0}^{\Nb-1}\,det({\bf 1}+a\hM(k))\bigg\rb^{1/2}
det({\bf 1}-e^{\alpha\beta}\,\hV(\Nb)) \label{31} 
\ee
where $\hM(k):=\frac{r'}{2a'}\hat{\Delta}(k)+m$ (i.e. the scalar part of $\hD_{space}(k)+m$).
The linear maps $\hcU^{(r)}(\Nb/2)$ on $W\oplus W$ and $\hV(\Nb)$ on $W$ in 
(\ref{22})--(\ref{31}) are defined as follows.
Exploiting the periodicity of the link variables to define $d_j^{(r)}(k)$ for all
$k\in{\bf Z}$, periodic under $k\to k+\Nb\,$, and thereby define $\hD^{(r)}\hpsi(k)$
for all $k\in{\bf Z}\,$, we consider the equation $\hD^{(r)}\hpsi(k)=0$
(no periodicity requirement on $\hpsi(k)$).
In the $r\ne1$ case, since the $d_{\pm1}^{(r)}(k)$'s are invertible, it is clear from
(\ref{15}) that solutions $\hpsi(k)$ are specified by two initial values. Thus the solution
space is isomorphic to $W\oplus W$. Setting $\hpsi_1(n)=\hpsi(2n)$ and 
$\hpsi_2(n)=\hpsi(2n+1)$ the solutions are determined from their initial values via 
an evolution operator: $\Big({\hpsi_1(n) \atop \hpsi_2(n)}\Big)=\hcU^{(r)}(n)
\Big({\hpsi_1(0) \atop \hpsi_2(0)}\Big)$. 
The operator $\hcU^{(r)}(\Nb/2)$ appearing in (\ref{22}) can also be characterised
as follows: Due to the $\Nb$--periodicity of the $d_j^{(r)}(k)$'s in (\ref{15})
there is a linear map on the solution space defined by $\hpsi(k)\mapsto\hpsi(k+\Nb)$, 
or, equivalently, $(\hpsi_1(n),\hpsi_2(n))\mapsto(\hpsi_1(n+\Nb/2),\hpsi_2(n+\Nb/2))$.
This map coincides with $\hcU^{(r)}(\Nb/2)$ when the solution space is identified
with $W\oplus W$.

In the $r=1$ case, solutions to $\hD^{(1)}\hpsi(k)=0$ are determined by just a single 
initial value; this is connected with the non-invertibility of $d_{\pm1}^{(1)}(k)$
and can be seen, e.g., from the expression (\ref{40}) below. Thus the solution space in 
this case is isomorphic to $W$. The evolution operator $\hV(k)$ determines solutions 
from their initial value through $\hpsi(k)=\hV(k)\hpsi(0)$. The $\hV(\Nb)$ in 
(\ref{31}) can be alternatively characterised as the linear map on the solution space
which maps $\hpsi(k)\mapsto\hpsi(k+\Nb)$.

Finite difference approximations to differential operators in one variable and their
determinants have been studied in \cite{BFK,Forman} and we are going to use a convergence 
result from there. 
In the setting of \cite{BFK,Forman}, specialising to 1st order differential operator,
the operator $L$ and its finite difference approximation $\hL$ have the forms
\be
L=L_1(\t)\frac{d}{d\t}+L_0(\t)\quad,\qquad\ \hL=\hL_1(k)\,\frac{1}{a}\,\partial+\hL_0(k)
\label{x1}
\ee
($\t\in{\bf R}\,$, $k\in{\bf Z}$) $\partial\in\{\partial^+,\partial^-\}\,$,
$\partial^+\hpsi(k)=\hpsi(k+1)-\hpsi(k)\,$, $\partial^-\hpsi(k)=\hpsi(k)-\hpsi(k-1)\,$,
with $L_j(\t),\hL_j(k):W\to W$ being periodic under $\t\to\t+\beta\,$, $k\to k+\Nb\,$,
respectively, and 
\be
\hL_j(k)=L_j(ka)+O(a)\qquad\ \mbox{($j=0,1$)}.
\label{x3}
\ee
Then the solutions to $L\Psi(\t)=0$ and $\hL\hpsi(k)=0$ are both determined by a single
initial value, so the solution spaces in both cases are isomorphic to $W$. Solutions $\hpsi$
approximate solutions $\Psi\,$, i.e. if $\hpsi(0)=\Psi(0)$ then 
$\hpsi(k)\approx\Psi(ka)$ for small $a$. Consequently the evolution operator 
$\hcU(k)$ for $\hL\hpsi=0$ approximates the evolution operator $\U(\t)$ for $L\Psi=0$.
(Explicitly, $\U(\t)=Te^{-\int_0^{\t}L_1(t)^{-1}L_0(t)\,dt}$.)
In particular one has the following (cf. \S3 of \cite{Forman}): 

\vspace{1ex}

\noindent {\em Convergence Theorem}: 
$\hcU(\Nb)\to\U(\beta)$ for $a\to0$ with $a\Nb=\beta$ held fixed.

\vspace{1ex}

\noindent An obvious variant of this which we will make use of is the following. 
If $p$ is a multiple of
$\Nb$ and $\hL=\hL_1\,\frac{1}{pa}\,\partial+\hL_0$ with the $\hL_j(k)$'s periodic under
$k\to k+\Nb/p$ and satisfying
$\hL_j(k)=L_j(kpa)+O(a)\,$ ($j=0,1$) then $\hU(\Nb/p)\to\U(\beta)$
for $a\to0$. Furthermore, if $W$ is replaced by $W_1\oplus W_2$ in the preceding 
then the convergence theorem continues to hold when $\partial$ is replaced by
$\Big({\partial \atop 0}\;{0 \atop \tilde{\partial}}\Big)$ with
$\partial,\tilde{\partial}\in\{\partial^+\,,\,\partial^-\}$.

In order to apply the convergence theorem to evaluate the $a\to0$ limits of 
(\ref{22})--(\ref{31}) we need to rewrite $\hD^{(r)}$ in the form of $\hL$ in (\ref{x1}),
or its aforementioned variant. We have only been able to do this 
in the $r=0$ and $r=1$ cases. The problem of evaluating
$\lim_{a\to0}\,detD_{\alpha}^{(r)}$ in the general $r$ case therefore remains for 
future work; new techniques beyond those of \cite{BFK,Forman} may be required for this.
In the $r=1$ case we specialise to a $\gamma$-representation where
$\g4=\Big({{\bf 1} \atop 0}\;{0 \atop -{\bf 1}}\Big)$ and decompose $W=W_+\oplus W_-$
so that $\g4=\pm1$ on $W_{\pm}$. Then, in terms of this decomposition,
\be
\hD^{(1)}=\hL_1^{(1)}\,\frac{1}{a}\,\bigg(\,{\partial^- \atop 0}\ {0 \atop \partial^+}
\bigg)+\hL_0^{(1)}
\label{40}
\ee
where
\be
\hL_1^{(1)}(k)&=&\g4\,\bigg(\,{U_4((k-1)a)^{-1} \atop 0}
\ {0 \atop U_4(ka)}\bigg) \label{41} \\
\hL_0^{(1)}(k)&=&\g4\,\bigg(\,{\frac{1}{a}(1-U_4((k-1)a)^{-1}) \atop 0}
\ {0 \atop \frac{1}{a}(U_4(ka)-1)}\bigg)+D_{space}(ka)+m \qquad\  
\label{42} 
\ee
Clearly $\hL_j^{(1)}(k)$ is periodic under $k\to k+\Nb$ and 
\be
\hL_1^{(1)}(k)=\g4+O(a)\quad,\qquad\ \hL_0^{(1)}(k)=\g4\,A_4(ka)+D_{space}(ka)+m+O(a)\,.
\label{47}
\ee
The convergence theorem now gives $\lim_{a\to0}\ \hV(\Nb)=\V(\beta)$
where $\V(\t)$, acting on $W$, is the evolution operator for $D\Psi(\t)=0$ with 
$D$ being the Dirac operator (\ref{8}) of the continuous time---lattice space setting.
Using this and noting $det({\bf 1}+a\hM(k))=e^{\,aTr\,\hM(k)}+O(a^2)$
the $a\to0$ limit of (\ref{31}) is now obtained:
\be
\lim_{a\to0}\ detD_{\alpha}^{(1)}=\Big(\frac{1}{a}\Big)^{N\Nb}
\,e^{-\alpha\beta N/2+\frac{1}{2}\int_0^{\beta}Tr\,M(\t)\,d\t}
\;det({\bf 1}-e^{\alpha\beta}\,\V(\beta))
\label{51}
\ee
The gauge field-independent factor $1/a^{N\Nb}\,$, which diverges in the $a\to0$ limit,
is physically inconsequential; it can at most give rise to an overall constant shift in the
calculation of certain physical quantities (such as the energy density in finite 
temperature QCD).

An application of the zeta-regularised determinant formula for differential operators
in one variable, Theorem 1 of \cite{BFK}, leads to an expression for $det_{\zeta}D_{\alpha}$ 
which coincides with (\ref{51}) without the $1/a^{N\Nb}$ factor and with $M(\t)$ replaced
by $\pm M(\t)$ (the details of this are given in \cite{DA(det)}).
The sign $\pm$ depends on the choice of cut in the complex plane used
to define the zeta-determinant. Choosing this so that the sign is ``$+$'' we then have 
$\lim_{a\to0}\ a^{N\Nb}\;detD_{\alpha}^{(1)}=det_{\zeta}D_{\alpha}$
which establishes the first part of (\ref{13}).

In the $r=0$ case, with $\hpsi$ represented by $(\hpsi_1(n),\hpsi_2(n))$ as before, we have
\be
\hD^{(0)}=\hL_1^{(0)}\,\frac{1}{2a}\,\bigg(\,{\partial^+ \atop 0}\ {0 \atop \partial^-}
\bigg)+\hL_0^{(0)}
\label{35}
\ee
where 
\be
\hL_1^{(0)}(n)&=&\bigg(\,{0 \atop \hK_1(n)}\ {\hJ_1(n) \atop 0}\bigg)
\quad,\quad\ \hL_0^{(0)}(n)=\bigg(\,{D_{space}(2na)+m \atop \hJ_0(n)}\ 
{\hK_0(n) \atop D_{space}((2n+1)a)+m}\bigg)\label{36}\\
\hJ_1(n)&=&\g4\,U_4((2n-1)a)^{-1}\quad,\quad\ \hK_1(n)=\g4\,U_4((2n+1)a) \label{37} \\
\hJ_0(n)&=&\g4\,\frac{1}{2a}\,\Big(\,U_4(2na)-U_4((2n-1)a)^{-1}\Big)\quad,\quad\ 
\hK_0(n)=\g4\,\frac{1}{2a}\,\Big(\,U_4((2n+1)a)-U_4(2na)^{-1}\Big)
\label{39}
\ee
Clearly $\hL_j^{(0)}(n)$ is periodic under $n\to n+\Nb/2$ and
\be
\hL_1^{(0)}(n)=\bigg(\,{0 \atop \g4}\ {\g4 \atop 0}\bigg)+O(a)\quad,\quad\ 
\hL_0^{(0)}(n)=\bigg(\,{D_{space}(2na)+m \atop \g4\,A_4(2na)}
\ {\g4\,A_4(2na) \atop D_{space}(2na)+m}\bigg)+O(a)\,.
\label{53}
\ee
The convergence theorem then gives $\lim_{a\to0}\ \hU(\Nb/2)=\U(\beta)$
where $\U(\beta)$, acting on $W\oplus W$, is the evolution operator for 
$\widetilde{D}\,\Big({\Psi_1 \atop \Psi_2}\Big)(\t)=0$ with
\be
\widetilde{D}=\bigg(\,{D_{space}(\t)+m \atop \g4\,(\frac{d}{d\t}+A_4(\t))}\ \
{\g4\,(\frac{d}{d\t}+A_4(\t)) \atop D_{space}(\t)+m}\bigg)\,.
\label{53a}
\ee
Introducing $\O=\Big({{\bf 1} \atop -{\bf 1}}\ {{\bf 1} \atop {\bf 1}}\Big)$ on
$W\oplus W$, we find after a little calculation
$\O^{-1}\,\widetilde{D}\,\O=\Big(\,{(\g4\gg5)^{-1}\,D\,(\g4\gg5) \atop 0}\ \ 
{ 0 \atop D}\Big)$,
where $D$ is the Dirac operator (\ref{8}). It follows that
$\U(\beta)=\O^{-1}\,\Big(\,{(\g4\gg5)^{-1}\,\V(\beta)\,(\g4\gg5) \atop 0}
\ \ {0 \atop \V(\beta)}\Big)\,\O$.
Using this, the $a\to0$ limit of (\ref{22}) is now obtained:
\be
\lim_{a\to0}\ detD_{\alpha}^{(0)}=\Big(\,\frac{1}{2a}\Big)^{N\Nb}\;e^{-\alpha\beta N}
\;det^2({\bf 1}-e^{\alpha\beta}\,\V(\beta))
\label{58}
\ee
Again there is a physically inconsequential, divergent factor, $(1/2a)^{N\Nb}$.
Comparing (\ref{58}) with (\ref{51}), and noting that 
$e^{\frac{1}{2}\int_0^{\beta}Tr\,M(\t)\,d\t}=e^{m\beta N/2}\;
e^{\frac{1}{2}\int_0^{\beta}Tr\,(\frac{r'}{2a'}\Delta_{space}(\t))\,d\t}$ where
$e^{m\beta N/2}$ is a p.i.f., we obtain the claimed result (\ref{12}).
Furthermore, an application of Theorem 1 of \cite{BFK} gives 
$det_{\zeta}(\gamma_4D_{\alpha})
=e^{-(1\mp1)\beta\alpha N/2}\,det({\bf 1}-e^{\alpha\beta}\,\V(\beta))$ where the sign 
$\mp$ again depends on the choice of cut in the complex plane \cite{DA(det)}.
It follows that
$\lim_{a\to0}\ (2a)^{N\Nb}\;detD_{\alpha}^{(0)}
=e^{\pm\beta\alpha N}\,det_{\zeta}(\gamma_4D_{\alpha})^2$. Since $e^{\pm\beta\alpha N}$
is a p.i.f. this establishes the second part of (\ref{13}).

In the free fermion case, $Z=detD_{\alpha}^{(r)}$ with $\alpha=\mu+i\pi/\beta$ is the
partition function of a free lattice Dirac fermion gas at temperature $T=1/\beta$
and chemical potential $\mu$. After noting 
$det({\bf 1}-e^{\alpha\beta}\V(\beta))=det({\bf 1}+e^{\beta\mu}\,e^{-\beta H})$ where
$H=\g4\,(D_{space}+m)$ it is easy to check that physical quantities 
such as the mean energy 
$\la E\ra=-\frac{\partial}{\partial\beta}\,logZ$ (with $\beta\mu$ held constant)
calculated from the expressions in this paper are consistent with 
the standard results described, e.g., in \cite{Rothe}. 
One subtlety is that in the $r=1$ case $\lim_{a\to0}\;\la E\ra$ has an additional
shift of $-2\sum_{\bp}(\frac{r'}{2a'}\Delta_{space}(\bp)+m)$ due to the exponential factor 
in (\ref{51}), while such a shift does not occur in the $r=0$ case (cf. (\ref{58})).
This shift can also be seen from the standard calculation in \S18.11 of \cite{Rothe}:
There the computation is reduced to a weighted sum of the residues of a certain function
$g(z)/z$. In addition to the poles at $z=z_{\pm}$ discussed there, there is also a pole at
$z=0$ and its residue is readily seen to give the aforementioned shift in 
$\lim_{a\to0}\;\la E\ra$. In the $r=0$ case there is no pole at $z=0$ and hence no
corrsponding shift in that case.

We have considered only the simplest lattice fermion formulations in this paper, namely 
the naive and Wilson fermions (with the connection to staggered fermions due to the 
relation between the naive and staggered fermion determinants).
More recent formulations of major current interest
are the domain wall \cite{Kaplan-Shamir} and overlap \cite{Neu(PLB)} fermions,
as well as other chirally improved approaches to lattice fermions \cite{other}. 
It appears feasible to evaluate the continuous time limit of the domain wall fermion
determinant (with finite, discretised fifth dimension) along the same lines as here;  
this is currently under investigation. Doing this in the case of overlap fermions
seems more technically difficult though. We remark that the overlap fermion determinant
has previously been successfully tested (in both the chiral and vector cases)
against known continuum expressions in 2 dimensions \cite{overlap2D}.

I thank Pierre van Baal and Claude Bernard for useful discussions and feedback.
The author is supported by the European Commission, contract HPMF-CT-2002-01716.

\end{document}